# Communicability Betweenness in Complex Networks

Ernesto Estrada[1,3*], Desmond J. Higham[2] and Naomichi Hatano[3]

[1] *Institute of Complex Systems at Strathclyde*, Department of Physics and Department of Mathematics, University of Strathclyde, Glasgow G1 1XH, UK

[2] Department of Mathematics, University of Strathclyde, Glasgow G1 1XH, UK

[3] Institute of Industrial Science, University of Tokyo, Komaba, Meguro, 153-8505, Japan

---

[*] Corresponding author. E-mail: estrada66@yahoo.com

**Abstract**

Betweenness measures provide quantitative tools to pick out fine details from the massive amount of interaction data that is available from large complex networks. They allow us to study the extent to which a node takes part when information is passed around the network. Nodes with high betweenness may be regarded as key players that have a highly active role. At one extreme, betweenness has been defined by considering information passing only through the shortest paths between pairs of nodes. At the other extreme, an alternative type of betweenness has been defined by considering all possible walks of any length. In this work, we propose a betweenness measure that lies between these two opposing viewpoints. We allow information to pass through all possible routes, but introduce a scaling so that longer walks carry less importance. This new definition shares a similar philosophy to that of communicability for pairs of nodes in a network, which was introduced by Estrada and Hatano (Phys. Rev. E 77 (2008) 036111). Having defined this new communicability betweenness measure, we show that it can be characterized neatly in terms of the exponential of the adjacency matrix. We also show that this measure is closely related to a Fréchet derivative of the matrix exponential. This allows us to conclude that it also describes network sensitivity when the edges of a given node are subject to infinitesimally small perturbations. Using illustrative synthetic and real life networks, we show that the new betweenness measure behaves differently to existing versions, and in particular we show that it recovers meaningful biological information from a protein-protein interaction network.

**Introduction**

Characterizing local properties of complex networks is important both in theory and practice [1]. While at the global scale we can summarize the overall architecture [2-4], at the local scale we can detect which nodes are the most relevant for the organization and functioning of a network [5-8]. These local measures are commonly named centrality indices [9] and have proved of great value in analyzing the role played by individuals in social networks [10]. We are now in a new era for the study of complex networks, which is characterized by the availability of large amounts of data representing biological, ecological, social, technological and infrastructural systems, among others [2-4]. Centrality measures have therefore become an important tool in identifying essential proteins [11-13], keystone species [14, 15], informational hubs or vulnerable infrastructures [16].

Centrality indices can be motivated by looking at the ability of a node to communicate directly with other nodes, or to its closeness at many other nodes or pairs of nodes or, more generally, by studying how frequently the node plays the role of intermediary when messages are passed around the network [17]. These ideas have given rise to some well-known centrality measures such as degree centrality (*DC*), closeness centrality (*CC*), and betweenness centrality (*BC*) [9, 18]. Other centrality measures, known as eigenvector centrality (*EC*), information centrality (*IC*) and the subgraph centrality (SC) were developed by Bonacich [19, 20], Stephenson and Zelen [21] and Estrada and Rodríguez-Velázquez [22, 23], respectively.

The *betweenness centrality*, *BC*, is defined as the fraction of shortest paths going through a given node [18]. It can be regarded as a measure of the "importance" of a node as a controller of the information which is flowing between the other nodes in the network. This measure assumes that all the information flowing through the network is going from one site to another following only the shortest (or geodesic) paths. However, it may be argued that most of the "information" is likely to flow through non-shortest paths. To account for this Freeman et al. [24] introduced the concept of *flow betweenness centrality* (FBC), which includes contributions from both shortest paths and non-shortest paths in the graph. However, Newman showed that the flow betweenness can give counterintuitive results in some cases [25], and offered the alternative concept of *random-walk betweenness centrality* (RWBC) [25]. The first two kinds of betweenness can be considered as mathematical ones while the third is a sort of statistically defined centrality.

The random-walk betweenness centrality of a node $r$ is equal to the number of times that a random walk starting and ending at two other nodes $p$ and $q$ passes through the node $r$ along the way, averaged over all $p$ and $q$. This centrality measure is based on a diametrically opposite premise to that of shortest-path betweenness [25]. In the shortest-path betweenness the "information" which travels from $p$ to $q$ through $r$ "knows" what is the most effective way to reach its target. In the Newman betweenness it is assumed that the "information" does not know anything about the best route to arrive at the target. In Newman's betweenness [25] we travel from $p$ to $q$ following links at random. Suppose that on the way node $r$ is visited many times. This indicates that node $r$ plays an important role in the transmission of information between $p$ and $q$. Averaging over all nodes $p$ and $q$ produces an overall centrality measure for node $r$.

The two approaches, shortest-path and random-walk betweenness, are at opposite ends of a spectrum, but each has a valid role in accounting for processes that can take place in the network. However, if they are at opposite ends of the betweenness scale it is interesting to ask what is in the middle. We believe that the answer is a betweenness centrality based on all walks connecting two nodes $p$ and $q$ that pass through $r$ weighted by path length; in other words, a betweenness centrality that accounts for the shortest path connecting two nodes, but also acknowledges the existence of other (non-shortest) paths, giving them less significance. This can be motivated by the idea is that the information travelling from $p$ to $q$ through $r$ knows the best way for arriving at the target but is also willing to make use of longer paths in order to deliver the information. The idea of using a weighted sum over walks of all possible lengths has been used successfully to define communicability between pairs of nodes [26]. Here we are extending the idea by looking at weighted all-walks passing through a third node. In the next section we show how to characterise and compute communicability betweenness via the exponential of an adjacency matrix. We then show some examples of the practical application of this measure in artificial and real-world networks.

**Preliminaries**

The main objective of the current work is to find a betweenness centrality measure that accounts for all the traffic which is carried through a particular node. This includes not only the shortest paths but every walk connecting the nodes $p$ and $q$ that pass through $r$. In this context we start by considering the communicability between a pair of nodes $p$ and $q$ [26].

The *communicability* between a pair of nodes in a network is usually considered as taking place through the shortest path connecting both nodes. However, it is known that communication between a pair of nodes in a network does not always take place through the shortest paths but it can follow other non-optimal walks. In a previous work we have introduced a communicability measure that fulfils these requirements [26]. If $P_{pq}^{(s)}$ is the number of shortest paths between distinct nodes $p$ and $q$ having length $s$ and $W_{pq}^{(k)}$ is the number of walks connecting $p$ and $q$ of length $k > s$, we may consider the quantity

$$G_{pq} = \frac{1}{s!} P_{pq}^{(s)} + \sum_{k>s} \frac{1}{k!} W_{pq}^{(k)}, \tag{1}$$

which was introduced in [26] to measure the communicability between nodes $p$ and $q$. This expression can be written as the sum of the $p, q$ entry of the different powers of the adjacency matrix

$$G_{pq} = \sum_{k=0}^{\infty} \frac{\left(A^k\right)_{pq}}{k!},$$
(2)

which converges to,

$$G_{pq} = \left(e^A\right)_{pq}.$$
(3)

There are several ways of weighting the walks according to their lengths. Some of them are now under study by two of the current authors and will be published elsewhere. The weighting used in (2) allows a physical interpretation of the communicability by considering a continuous-time quantum walk on the network. Considering this approach, the communicability between nodes $p$ and $q$ in the network represents the probability that a particle starting from the node $p$ ends up at the node $q$ after wandering on the complex network due to the thermal fluctuation. By regarding the thermal fluctuation as some form of random noise, we can identify the particle as an information carrier in a society or a needle in a drug-user network.

Our objective here is to use this communicability function to account for the betweenness centrality of a node in a complex network. It is clear that there are several ways for accounting such a property of a node. As we have already remarked the shortest-path and the random-walk betweenness are two extreme cases. Another possibility is to consider the number of paths or the number of self-avoiding walks between any two nodes [27]. Here, however, we consider the use of the number of walks connecting every pair of nodes as the basis for the new centrality measure. By using this approach we are able to define the betweenness centrality on the basis of the communicability function, which can be physically interpreted as the Green's function of the complex network.

Consequently, in full analogy with (1) we now let $G_{prq}$ be the corresponding weighted sum where we only consider walks that involve node $r$. The new concept of *communicability betweenness centrality* (CBC) of node $r$ then takes the form

$$\omega_r = \frac{1}{C} \sum_p \sum_q \frac{G_{prq}}{G_{pq}}, \qquad p \neq q, p \neq r, q \neq r,$$ (4)

where $C = (n-1)^2 - (n-1)$ is a normalization factor equal to the number of terms in the sum, so that $\omega_r$ takes values between zero and one. We show in the next section that this quantity can be neatly characterised in terms of the matrix exponential.

**Communicability betweenness**

Let $G = (V, E)$ be a simple graph with $n = |V|$ nodes and $m = |E|$ links, and let $\mathbf{A}$ denote its adjacency matrix. Let $G(r) = (V, E')$ be the graph resulting from removing all edges connected to the node $r \in V$, but not the node itself. The adjacency matrix for $G(r)$ may be written $\mathbf{A} + \mathbf{E}(r)$, where $\mathbf{E}(r)$ has nonzeros only in row and column $r$, and in this row and column has -1 wherever $\mathbf{A}$ has +1. For simplicity, we will often write $\mathbf{E}$ instead of $\mathbf{E}(r)$.

Analogously to (3), $G_{prq}$ may be written

$$G_{prq} = \left(e^{\mathbf{A}}\right)_{pq} - \left(e^{\mathbf{A}+\mathbf{E}(r)}\right)_{pq},$$ (5)

so in (4) we have

$$\omega_r = \frac{1}{C} \sum_p \sum_q \frac{G_{prq}}{G_{pq}} = \frac{1}{C} \sum_p \sum_q \frac{\left(e^{\mathbf{A}}\right)_{pq} - \left(e^{\mathbf{A}+\mathbf{E}(r)}\right)_{pq}}{\left(e^{\mathbf{A}}\right)_{pq}}, \quad p \neq q, p \neq r, q \neq r,$$ (6)

which gives a computable characterisation of CBC. We consider connected graphs only, for which $\left(e^{\mathbf{A}}\right)_{pq}$ is nonzero.

It is straightforward to realise that the proportion of communicabilities defining the betweenness (see Eq. (6)) is bounded between zero and one,

$$0 \leq \frac{\left(e^{A}\right)_{pq} - \left(e^{A+E(r)}\right)_{pq}}{\left(e^{A}\right)_{pq}} \leq 1.$$ (7)

The lower bound is obtained when $\left(e^{\mathbf{A}}\right)_{pq} = \left(e^{\mathbf{A}+\mathbf{E}(r)}\right)_{pq}$, which indicates that no walks connecting the nodes $p$ and $q$ pass through the node $r$. This cannot happen for a connected graph, but we will have $\left(e^{\mathbf{A}}\right)_{pq} \cong \left(e^{\mathbf{A}+\mathbf{E}(r)}\right)_{pq}$ when the walks that connect $p$ and $q$ through $r$ are of very long length.

On the other hand, the upper bound is reached for a graph in which all walks connecting the nodes $p$ and $q$ pass through the node $r$. This situation coincides with the selection of the node $r$ as the one with degree $n-1$ in a star with $n$ nodes, $S_n$. We recall that $S_n$ is the graph having one node with degree $n-1$ (central node) and $n-1$ nodes of degree 1. In this case if we remove the links incident with node $r$ we obtain the fully disconnected graph. Consequently, $\left(e^{\mathbf{A}+\mathbf{E}(r)}\right)_{pq} = 0$ for every pair of the nodes with degree 1 in the star and we obtain the upper bound for the expression (7).

If we consider now the expression for the communicability betweenness for a node $r$ in any graph we see that

$$\omega_r = \frac{1}{C}\sum_p \sum_q \frac{\left(e^A\right)_{pq} - \left(e^{A+E(r)}\right)_{pq}}{\left(e^A\right)_{pq}} \le \frac{(n-1)(n-2)}{C}, \tag{8}$$

which for the normalization constant selected in this work gives $\omega_r = 1$ for the central node of the star graph. The upper bound for the communicability betweenness is obtained for the central node of $S_n$ for which we have the maximum value of (7) for all pairs of non-hub nodes in the graph. This is a desired property of a betweenness measure; optimal centrality should correspond to the central hub of a star, through which all the communication between the other pairs of nodes passes. It is also clear from the construction that $\omega_r$ increases in any graph if node $r$ is given an extra edge.

Although the matrix exponential cannot be computed in finite time, a high-accuracy approximation can be obtained from an algorithm that is guaranteed to terminate. The scaling and squaring method for exponentiating a matrix in [28] (Algorithm 10.20) has an operation count that is a modest multiple of $n^3$, so that the overall cost of computing $\omega_r$ in (5) for all nodes in the graph scales like $n^4$. However, if the network is sparse then it will typically not be necessary to store or compute with a full $n \times n$ matrix. The series expansion (2) could be terminated after a predetermined number of terms, so that walks of sufficiently long length are ignored. Alternatively, existing high quality sparse matrix software, for example, the function *EIGS* in MATLAB, could be used to compute the dominant eigenvalues and corresponding eigenvectors and a truncated spectral formulation of the matrix exponential could then be exploited. In this case the efficiency gain would be highly dependent on the sparsity structure.

**Communicability betweenness and the Fréchet derivative**

We see from (5) that CBC depends on the relative componentwise changes that occur in the matrix exponential when a specific row/column is set to zero. Such a change represents an $O(1)$ perturbation. We will show now that this quantity is closely related to the sensitivity of the matrix exponential to small changes in the same direction, which gives another interpretation of the CBC measure.

The effect of making an infinitesimal perturbation to the adjacency matrix exponential in the direction $\mathbf{E}$ is given by the directional or Fréchet derivative,

$$D_{E(r)}(A) = \lim_{h \to 0} \frac{1}{h} \left[ e^{(A + hE(r))} - e^{A} \right],$$  (9)

which may be, in different contexts, called the Gáteaux derivative or a linear response [28-30]. The Fréchet derivative can also be expressed as a Taylor series as follows [28]

$$D_{E(r)}(A) = \sum_{k=1}^{\infty} \frac{1}{k!} \sum_{j=1}^{k} A^{j-1} E(r) A^{k-j} \quad .$$  (10)

We now interpret the right-hand side of (10) in terms of walks through the network. As before we consider a pair of distinct nodes, $p$ and $q$ that are both different from $r$. We first note that $(\mathbf{AE})_{pq} = -1$ if there is a walk (a path in this case) of the form $p \to r \to q$ and $(\mathbf{AE})_{pq} = 0$ otherwise. Similarly, $(\mathbf{EA})_{pq} = -1$ if there is a walk of the form $p \to r \to q$ and $(\mathbf{EA})_{pq} = 0$ otherwise. So, we see that the $k = 1$ term of the expansion in (10), $(\mathbf{AE} + \mathbf{EA})_{pq}$, contributes $-2$ if there is a walk of length two that connects nodes $p$ and $q$ via node $r$, and contributes $0$ otherwise. Similarly, it may be shown that the $k = 3$ term, $(\mathbf{A}^2\mathbf{E} + \mathbf{AEA} + \mathbf{EA}^2)_{pq}$, contributes $-2$ times the number of walks of length three from $p$ to $q$ that involve $r$. For the case $k \geq 3$, however, the relationship is not quite so simple. These terms are studied in the Appendix and our overall conclusion is that the $(p, q)$ element of the Fréchet derivative (9), when multiplied by $-1$, is *at least twice as big* as the weighted number of walks from $p$ to $q$ that involve $r$. It is generally not a lot more than twice as big, because the extra amount comes from walks that involve node $r$ more than once—these will generally be long and hence will have small weights.

To illustrate this idea, we constructed a network with 8 nodes where 19 edges were assigned arbitrarily. Fig. 1 illustrates the adjacency matrix. In this case the ratio of the $(p, q)$th component of the Fréchet derivate in (9) to the quantity $G_{prq}$ in (5), as we vary $p$ to $q$ and $r$, ranges from $-2.08$ to $-2.96$.

**Insert Fig. 1 about here.**

Overall, we have established a close relation between the limiting $h \to 0$ sensitivity in (9) and the $h = 1$ analogue. We emphasize that this relationship hinges on the special nature of the direction $\mathbf{E}(r)$ in which the adjacency matrix is perturbed.

**Analysis of the communicability betweenness**

We test here three examples that Newman [25] used to introduce the random-walk betweenness, plus an extra illustrative example. We first consider the two graphs represented in Fig. 2. These were selected by Newman [25] as prototypes of networks in which there are some nodes, like C, which are not intermediaries in many shortest paths, but may play an important role in the flow of information between distinct communities. The betweenness measures for these networks is shown in Table 1. We see that for network 1, CBC ranks node X higher than node C, unlike RWBC. This can be explained by the fact that node C is never an intermediary in a non-trivial shortest length walk, whereas node X is important for walks within its own clique. For network 2, node C plays a more significant role in short walks and CBC ranks C only slightly lower than A.

Network 3, shown in Fig. 3, illustrates that CBC and RWBC can differ dramatically. Here, we have two large communities, each consisting of a complete graph of 12 nodes. These communities are joined by a short chain through the single node A and also by a long chain of 8 nodes that has node B at one end and node C in the middle. We see in the Fig. 3 that while both measures pick out node A as the most central, the RWBC version ranks node B second and much prefers node C to node D while CBC ranks D second and, most notably, attaches very little significance to node C. This can be explained by the fact that node C is typically involved in relatively long walks that are down-weighted by the CBC measure. This discrepancy between the two centrality measures becomes even more pronounced if the size of the complete subgraphs and the length of the connecting chain are increased. When node A is removed in the network in Fig. 3, CBC behaves like RWBC in placing node C ahead of node D.

The fourth example in this section is the network of intermarriages between prominent families in early 15[th] century Florence [31] (see Fig. 4). The four betweenness measures identify

Medici as the most central family. However, there are some differences in the ranking for the rest of the families. For this specific network the communicability betweenness is highly correlated to the random walk one ($r^2 = 0.964$) and in lower degree with shortest path ($r^2 = 0.935$) and flow ($r^2 = 0.844$) betweenness. The intercorrelation between the different measures is dependent on the specific networks we are studying and it should be noted that even when high correlations are observed there can be significant differences in the ranking of individual nodes. For instance, the random walk betweenness ranks the Strozzi family as the fourth most central only after Medici, Guadagni and Albizzi. However, communicability centrality ranks this family as the sixth most central after Medici, Guadagni, Albizzi, Ridolfi and Tornabuoni.

**Insert Fig. 4 about here.**

We also study the correlation existing between the various betweenness centralities and the degree centrality for the network of marriages in Florence. The lowest correlations are observed for the flow ($r^2 = 0.662$) and shortest path betweenness ($r^2 = 0.712$) and the largest one for the random walk ($r^2 = 0.906$). The communicability betweenness displays an intermediate correlation ($r^2 = 0.856$). Despite these correlations "there are usually a small number of vertices in a network for which betweenness and degree are very different, and betweenness is useful precisely in identifying these vertices" [25]. For instance, communicability betweenness is able to differentiate the four less central families in this network, which neither of the other betweennesses nor the degree are able to differentiate. The communicability betweenness ranks these families in the following decreasing order, Acciaiuoli > Lamberteschi > Ginor > Pazzi.

To investigate more deeply the correlations between the different betweenness measures as well as the degree centrality we study a series of real-world networks, including social, informational, biological, technological and ecological (see [32] for references about these networks). The names, number of nodes ($n$), links ($m$) and a brief description of these networks are given in Table 2.

We study not only the correlation coefficient but also the relative error for the prediction of the CBC based on the other measures. The relative prediction error, in percentage, is expressed as

$$RPE(\%) = 100 \frac{SEP(y)}{\langle y \rangle} \tag{11}$$

where $SEP(y)$ is the standard error of predicting the variable $y$ using a regression model of the type $y = ax + b$ and $\langle y \rangle$ is the average of the $y$-values. Here $y$ is the communicability betweenness centrality and $x$ is any of the other centrality measures, i.e., DC, BC, FBC and RWBC.

The largest correlations for CBC are degree centrality for which the average correlation coefficient is 0.866 (see Table 3). The correlation between degree and betweenness has been previously observed by Kitsak et al. to be dependent on the fractal nature of networks [33]. These authors have observed that the correlation is much weaker in fractal than in non-fractal networks [33]. However, even in this case the relative prediction error is high (51.4%), which indicates that both centrality measures account for different aspects of the structure of a network. For instance, there are cases like the network of Colorado Springs in which the correlation between both measures is very poor ($r^2 = 0.589$) and the RPE is higher than 100%. In other cases like in the transcription network of E. coli the correlation coefficient is high ($r^2 = 0.864$) but the relative error is almost 100% ($RPE = 95.8\%$). This is because there is a node with very high value of both degree and CBC which pulls the line to one extreme making the correlation coefficient very high. However, the dispersion for the points below this extreme is very high as indicated by RPE.

The relationships between the CBC and the other betweenness measures are poorer. For instance, the average correlation coefficients rank from 0.44 for the flow betweenness to 0.78 for the random walks betweenness. However, the relative prediction errors are very high, overpassing 60% in all cases and reaching 102.8 for the FBC. This clearly indicates that the CBC is an independent measure of node betweenness, which accounts for certain topological and structural characteristics of nodes not accounted for either of the other measures of betweenness or the degree centrality. In the next section we will analyze in more detail how these differences are manifested in one particular complex network.

**Betweenness in a protein interaction network**

Here we study the protein-protein interaction network (PIN) of *E. coli* in more detail. This network is formed by 695 validated interactions between 230 proteins in the bacterium *E. coli*. Butland et al. [34] have observed that the most highly conserved proteins in this PIN are the most highly connected ones. That is, those proteins that have been evolutionarily conserved in many different bacterial genomes are those having the largest degree centrality in the *E. coli* PIN. The conservation of a protein was determined by Butland et al. [34] by searching for homologues in 148 different genomes using BLAST. These genomes include proteobacteria, bacilli, clostridia, actinobacteridae, mycoplasmas, chlamydiaceae, cyanobacteria, archae and eubacteria. These authors consider a protein to be conserved if it is detected in at least 125 of the 148 genomes. On the contrary, a protein is deemed to be non-conserved if it appears at most in 25 genomes.

We calculated the degree and betweenness centrality measures for every protein in the *E. coli* PIN. Then we ranked these proteins in decreasing order of their centrality and analysed how many proteins are conserved in the top 1/3, the middle 1/3 and the bottom 1/3. We calculated the proportion of conserved to non-conserved proteins for each centrality. The results are displayed in Fig. 5. We see that the random walk centrality performs the best in detecting the conserved proteins as it identifies more than 7 conserved proteins per each non-conserved one in the top 33% of the ranking. In addition it performs very well for the bottom 1/3 of the ranking having the lowest proportion of conserved to non-conserved proteins. The second best performance is displayed by the communicability betweenness, which identifies more than 5 conserved proteins per each non-conserved one in the top 1/3 of the ranking. It also displays the second best performance for the bottom 1/3 of the ranking. The worst results in this case are obtained for the flow betweenness, which identifies only 2 conserved proteins per each non-conserved one in the top 1/3 of the ranking.

**Insert Fig. 5 about here.**

In addition, Butland et al. [34] observed a positive correlation between the number of protein interactions, i.e., protein degree, and the number of genomes a homolog was detected in. The Pearson correlation coefficient for this relation is 0.24 as illustrated in Fig. 5 of the Supplementary material of that paper [34]. We have found here that the communicability betweenness displays the same correlation as the degree centrality with the number of genomes in which a protein is conserved, i.e., a Pearson correlation coefficient of 0.24. The random-walk betweenness displays a correlation coefficient of 0.22, the shortest-path betweenness a correlation coefficient of 0.15 and the flow betweenness a correlation coefficient of 0.11. To explore these effects further, we may consider the top 20 proteins according to the ranking introduced by the centrality measures. In Fig. 6 we illustrate the results obtained for this analysis, where it can be seen that the proteins ranked by the communicability betweenness display the largest average number of genomes in which such proteins appear.

**Insert Fig. 6 about here.**

The same situation is repeated when we select the top 30 or top 40 proteins according to each ranking. These results indicate that the communicability betweenness display the best performance in identifying proteins which are conserved in the largest number of genomes.

Classifying a protein as highly conserved if it appears in at least 95 of the 148 genomes, we found that both the random-walk and communicability betweenness display the best performance in identifying the most conserved proteins among the top 1/3 proteins ranked by these measures (see Fig. 7). They both identify 54 of the most highly conserved proteins from 76 possible ones, which is more than 70%. The second best performance is obtained for the degree centrality and the worst is again for the flow centrality. However, if we analyze the top 2/3 proteins ranked by each centrality measure, the communicability betweenness overtakes all the other centralities. It identifies 95 of the 152 most highly conserved proteins, followed by the degree and shortest-path betweenness which identifies 93 of such proteins. For the bottom 1/3 of the ranking the best performance is again observed for the communicability betweenness which left only 25 of the most highly conserved proteins among the less central ones. The worse performance in this case is observed for the random-walk betweenness which left 29 of the most highly conserved proteins among the less central ones.

**Insert Fig. 7 about here.**

**Conclusion**

One point that should be clear from this and other studies analysing centrality measures in complex networks is that there is not one centrality which is the best or worst. Centrality measures account for different aspects of the topological relevance of a node. Consequently, a centrality can be good for summarizing one network feature and poor for another. This emphasizes the importance of (a) providing theoretical justification and analysis, and (b) testing on real complex networks, in order to understand what these measures have to offer. We have shown here that the communicability betweenness is amenable to both approaches. On the one hand, the communicability betweenness, which measures the changes in the adjacency matrix exponential subject to a certain $O(1)$ perturbation, is closely related to the instantaneous rate of change. In other words, although communicability betweenness was motivated by looking at what happens when the edges for node $r$ are removed, it also measures the sensitivity of a node's communicability when its edges are subject to *infinitesimal* changes. (In the latter case, we move into the realm of graphs with weighted edges.). On the other hand, we have shown empirically that this centrality measure accounts for important topological characteristics of the nodes in real-world complex networks. Hence we believe that this new concept has value in the study of centrality in complex networks.

**Appendix**

For the general term of (10) where $k \geq 3$ there is an expression of the form,

$$\left(\mathbf{A}^{k-1}\mathbf{E}\right)_{pq} = \sum_{s_1=1}^{n}\sum_{s_2=1}^{n}\cdots\sum_{s_{k-1}=1}^{n} a_{ps_1} a_{s_1 s_2} a_{s_2 s_3} \cdots a_{s_{k-2} s_{k-1}} e_{k-1 q}\ .$$

We see that $\left(\mathbf{A}^{k-1}\mathbf{E}\right)_{pq} = -1$ if there is a walk

$$p \to s_1 \to s_2 \to s_3 \cdots \to s_{k-2} \to r \to q\ , \tag{A1}$$

for any nodes $s_1, s_2, \cdots, s_{k-2},\ $, such that successive nodes are distinct, and $\left(\mathbf{A}^{k-1}\mathbf{E}\right)_{pq} = 0$ otherwise.

For the general case,

$$\left(\mathbf{A}^{j-1}\mathbf{E}\mathbf{A}^{k-j}\right)_{pq} = \sum_{s_1=1}^{n}\sum_{s_2=1}^{n}\cdots\sum_{s_{k-1}=1}^{n} a_{ps_1} a_{s_1 s_2} a_{s_2 s_3} \cdots a_{s_{j-2} s_{j-1}} e_{s_{j-1} s_j} a_{s_j s_{j+1}} \cdots a_{s_{k-1} q}\ ,$$

we see that $\left(\mathbf{A}^{j-1}\mathbf{E}\mathbf{A}^{k-j}\right)_{pq} = -1$ if there is a walk

$$p \to s_1 \to s_2 \to s_3 \cdots \to s_{j-1} \to r \to s_{j+1} \to \cdots \to s_{k-1} \to q\ , \tag{A2}$$

or

$$p \to s_1 \to s_2 \to s_3 \cdots \to s_{j-2} \to r \to s_j \to s_{j+1} \to \cdots \to s_{k-1} \to q\ , \tag{A3}$$

and is zero otherwise.

Finally, since

$$\left(\mathbf{E}\mathbf{A}^{k-1}\right)_{pq} = \sum_{s_1=1}^{n}\sum_{s_2=1}^{n}\cdots\sum_{s_{k-1}=1}^{n} e_{ps_1} a_{s_1 s_2} a_{s_2 s_3}\cdots a_{s_{k-2}s_{k-1}} a_{s_{k-1}q}\,,$$

it follows that $\left(\mathbf{E}\mathbf{A}^{k-1}\right)_{pq} = -1$ if there is a walk

$$p \rightarrow r \rightarrow s_2 \rightarrow s_3 \rightarrow s_4 \cdots \rightarrow s_{k-1} \rightarrow q\,, \tag{A4}$$

and is zero otherwise.

Note that any walk appearing in (A1) also appears in the $j=k-1$ case of (A2), any walk appearing in the general $(j-1)$st instance of (A2) also appears in the $j$th instance of (A3) and any walk appearing in (A2) also appears in the $j=2$ case of (A3). This shows that, for general walks of length $k$, (i) each walk involving node $r$ is counted twice, and (ii) the count involves *multiplicity*: e.g., a walk that involves node $r$ twice counts as two walks (i.e. counts four times in total because of point (i)), and, generally, a walk that involves node $r$ a total of $d$ times will count as $d$.

To illustrate point (ii), the walk $p \rightarrow r \rightarrow 3 \rightarrow r \rightarrow q$ shows up in the form $p \rightarrow s_1 \rightarrow s_2 \rightarrow r \rightarrow q$ and in the form $p \rightarrow r \rightarrow s_2 \rightarrow s_3 \rightarrow q$ (and in each case is counted twice). This multiplicity effect is relatively minor because it only arises for walks of length greater than or equal to four. (Walks from $p$ to $q$ of length two or three cannot involve $r$ more than once.) Hence, because of the $k!$ scaling in (A2), the multiplicities get down-weighted.

**Acknowledgements**

EE thanks A. Emili and J. M. Peregrín-Alvarez for sharing the datasets of the *E. coli* PIN and some useful comments about conserved and essential proteins, as well as the support by the Royal Society of Edinburgh and the Edinburgh Mathematical Society during a visit to the Department of Mathematics, University of Strathclyde (March 2008), to the IIS, University of Tokyo for a fellowship as Research Visitor during April-June, 2008 and both institutions for warm hospitality.

DJH is supported by the Engineering and Physical Sciences Research Council of the UK under grant GR/S62383/01.

Table 1. Values of the different betweenness centrality measures for the nodes labeled in the graphs represented in the Fig. 2.

| Network | Node | Betweenness measure | | | |
|---|---|---|---|---|---|
| | | BC | FBC | RWBC | CBC |
| 1 | A | 0.636 | 0.631 | 0.670 | 0.657 |
| | C | 0.200 | 0.282 | 0.333 | 0.140 |
| | X | 0.200 | 0.068 | 0.269 | 0.216 |
| 2 | A | 0.265 | 0.269 | 0.321 | 0.166 |
| | C | 0.243 | 0.004 | 0.267 | 0.157 |
| | X | 0.125 | 0.024 | 0.194 | 0.128 |

Table 2. Description of the real-world complex networks studied in the current work.

| Name | $n$ | $m$ | description |
|---|---|---|---|
| Galesburg | 31 | 67 | friendship ties among 31 physicians |
| High Tech | 33 | 91 | friendship ties among the employees in a small hi-tech computer firm which sells, installs, and maintains computer systems |
| Prison | 67 | 142 | social network of inmates in prison who chose "What fellows on the tier are you closest friends with? |
| College | 32 | 96 | social network among college students in a course about leadership, where the students choose which three members they wanted to have in a committee |
| Zachary | 34 | 78 | a network of friendship between members of the Zachary karate club |
| ColoSpring | 324 | 347 | risk network of persons with HIV infection during its early epidemic phase in Colorado Springs, USA (sexual and injecting drugs partners) from 1985-1999 |
| Centrality | 118 | 613 | citation network of papers published in the field of Network Centrality |
| Small World | 233 | 994 | papers that cite S. Milgram's 1967 Psychology Today paper or use Small World in the title |
| Electronic1 | 122 | 189 | electronic sequential logic circuits parsed from the ISCAS89 benchmark set, where nodes represent |
| Electronic2 | 252 | 399 | logic gates and flip-flops |
| USAir97 | 332 | 2126 | the airport transportation network between airports in US in 1997 |
| PIN-Afulgidus | 32 | 36 | protein-protein interaction network of *A. fulgidus* |

| PIN-Bsubtilis | 84 | 98 | protein-protein interaction network of *B. subtilis* |
| PIN-E.coli | 230 | 695 | protein-protein interaction network of the bacterium *E. coli* |
| Neurons | 280 | 456 | the neuronal synaptic network of the nematode *C. elegans*, including all data except muscle cells and using all synaptic connections |
| Trans-Ecoli | 328 | 456 | the direct transcriptional regulation between operons in *Escherichia coli* |
| Bridge Brook | 75 | 542 | represents the pelagic species from the largest of a set of 50 New York Adirondack lake food webs |
| Canton Creek | 108 | 613 | invertebrates and algae in a tributary, surrounded by pasture, of the Taieri River in the South Island of New Zealand |
| El Verde | 156 | 1439 | the insects, spiders, birds, reptiles and amphibians in a rainforest in Puerto Rico; |
| Grassland | 75 | 113 | all vascular plants and all insects and trophic interactions found inside stems of plants collected from 24 sites distributed within England and Wales |
| Little Rock | 181 | 2318 | the pelagic and benthic species, particularly fishes, zooplankton, macroinvertebrates, and algae of the Little Rock Lake, Wisconsin, U.S |
| Reef Small | 50 | 503 | Caribbean coral reef ecosystem from the Puerto Rico-Virgin Island shelf comple |

Table 3. Statistical parameters for the correlations between the different centrality measures studied here for 22 real-world complex networks.

| Network | DC | | BC | | FBC | | RWBC | |
|---------|-------|---------|-------|---------|-------|---------|-------|---------|
| | $r^2$ | $RPE(\%)$ | $r^2$ | $RPE(\%)$ | $r^2$ | $RPE(\%)$ | $r^2$ | $RPE(\%)$ |
| Galesburg | 0.918 | 26.5 | 0.763 | 45.2 | 0.426 | 70.5 | 0.815 | 39.49 |
| HighTech | 0.969 | 17.3 | 0.644 | 58.5 | 0.294 | 82.6 | 0.754 | 47.50 |
| Prison | 0.893 | 31.8 | 0.851 | 36.3 | 0.202 | 85.4 | 0.862 | 33.49 |
| Social3 | 0.941 | 19.8 | 0.862 | 30.5 | 0.452 | 61.7 | 0.837 | 33.93 |
| Zackarias | 0.957 | 23.8 | 0.852 | 44.0 | 0.736 | 59.2 | 0.955 | 25.31 |
| ColoSpg | 0.589 | 146.3 | 0.990 | 17.0 | 0.607 | 147.1 | 0.941 | 16.30 |
| PIN-Afulgidus | 0.887 | 45.6 | 0.993 | 11.5 | 0.823 | 57.2 | 0.966 | 25.80 |
| PIN-Bsubtilis | 0.871 | 67.9 | 0.988 | 20.4 | 0.804 | 83.5 | 0.923 | 50.23 |
| PIN-Ecoli | 0.900 | 58.4 | 0.298 | 156.6 | 0.101 | 177.5 | 0.408 | 152.73 |
| Neurons | 0.828 | 71.4 | 0.722 | 90.0 | 0.413 | 131.3 | 0.711 | 98.96 |
| Trans-Ecoli | 0.864 | 95.8 | 0.953 | 56.2 | 0.701 | 141.6 | 0.903 | 77.66 |
| Centrality | 0.957 | 28.5 | 0.578 | 89.8 | 0.300 | 115.2 | 0.756 | 67.02 |
| SmallWorld | 0.871 | 68.8 | 0.503 | 134.4 | 0.543 | 129.1 | 0.645 | 115.64 |
| Electronic1 | 0.626 | 74.5 | 0.989 | 12.5 | 0.244 | 106.1 | 0.872 | 42.50 |
| Electronic2 | 0.526 | 108.1 | 0.995 | 11.7 | 0.127 | 152.2 | 0.807 | 79.57 |
| USAir97 | 0.940 | 51.1 | 0.350 | 167.4 | 0.204 | 186.2 | 0.561 | 132.25 |
| BridgeBrook | 0.882 | 31.7 | 0.501 | 65.0 | 0.297 | 77.3 | 0.689 | 52.19 |
| Canton | 0.960 | 19.8 | 0.536 | 67.7 | 0.448 | 74.1 | 0.771 | 48.60 |
| ElVerde | 0.957 | 25.1 | 0.482 | 86.1 | 0.327 | 98.4 | 0.643 | 72.08 |
| Grassland | 0.835 | 74.5 | 0.957 | 35.8 | 0.768 | 88.5 | 0.963 | 34.25 |
| LittleRock | 0.903 | 33.8 | 0.291 | 91.3 | 0.127 | 101.6 | 0.462 | 79.11 |
| ReefSmall | 0.976 | 9.5 | 0.619 | 38.2 | 0.689 | 34.4 | 0.860 | 22.79 |

Fig. 1. Adjacency matrix for a graph with 8 nodes and 19 edges used to illustrate the Fréchet derivative. Nonzeros are indicated by dots.

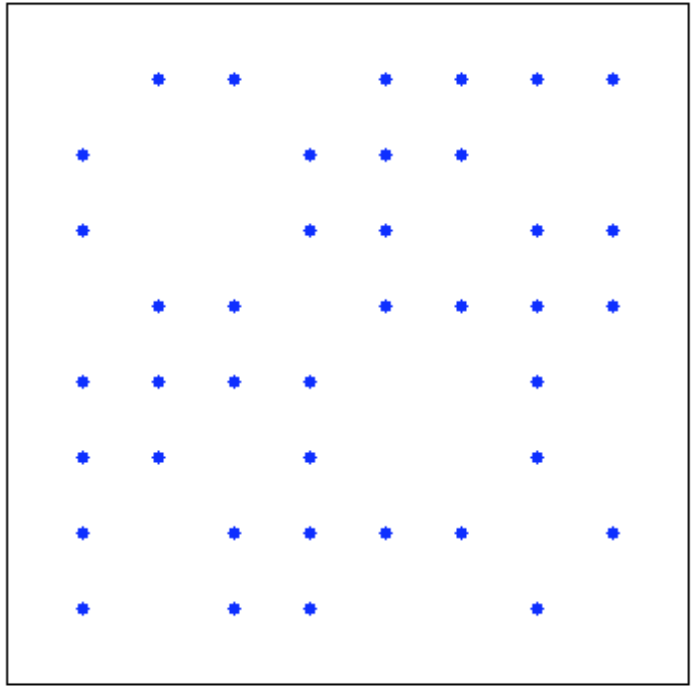

Fig. 2. Networks used by Newman [25] as examples for the analysis of the differences between centrality measures. The results obtained here including the communicability betweenness are given in Table 1.

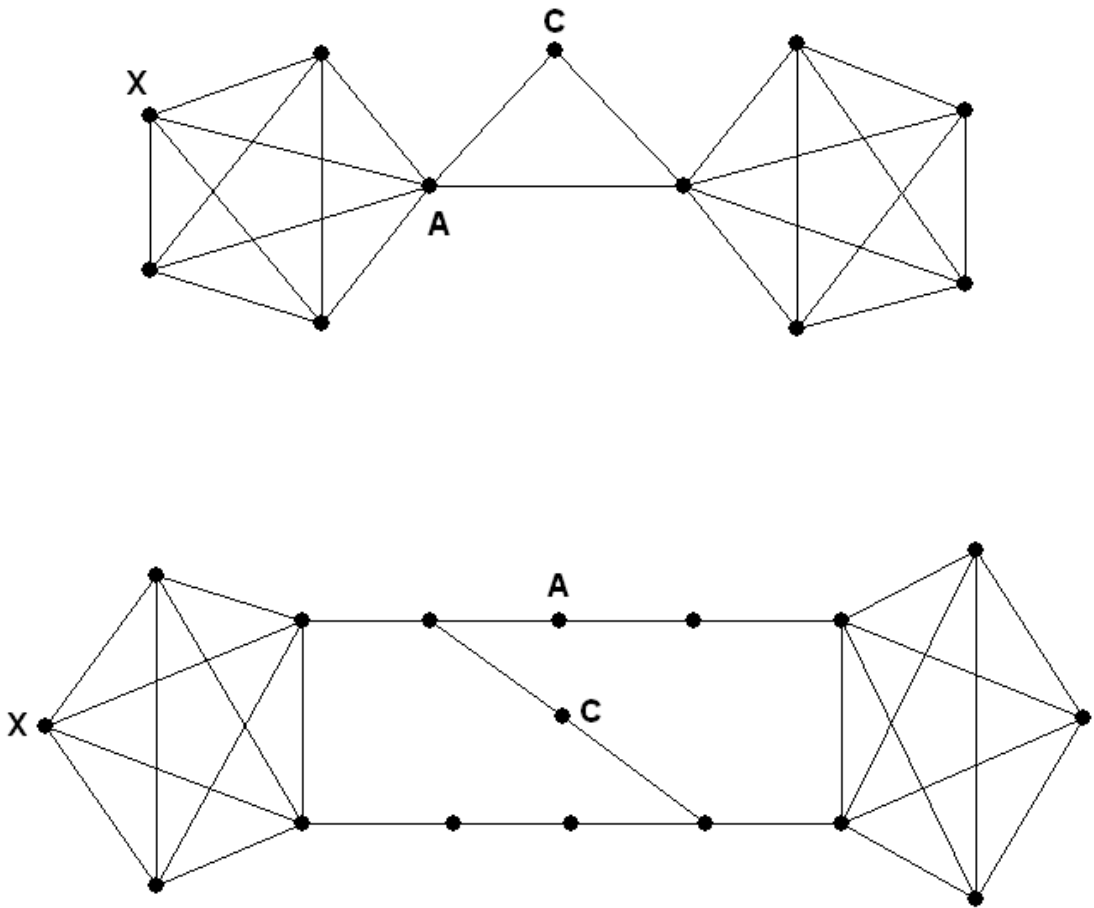

Fig. 3. Network where two large communities are connected through a single node and also through a longer chain.

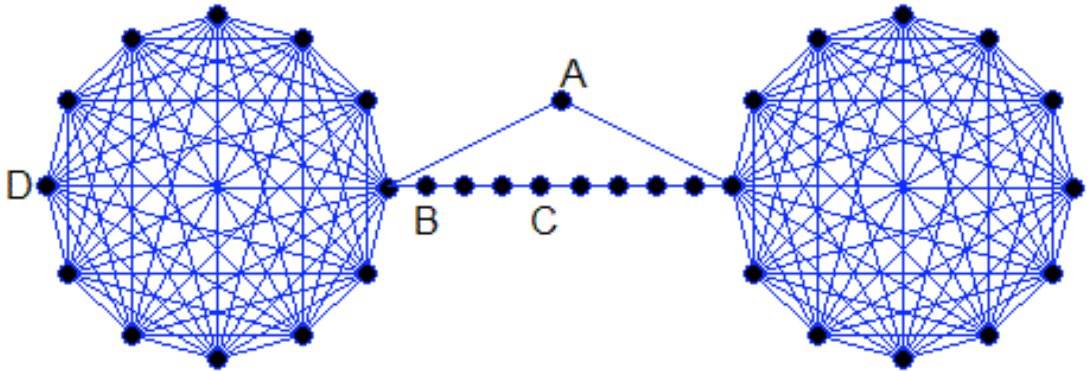

| Node | RWBC | CBC |
|------|-------|-------|
| A | 0.447 | 0.517 |
| B | 0.275 | 0.225 |
| C | 0.237 | 0.129 |
| D | 0.102 | 0.313 |

Fig. 4. Illustration of the network of intermarriages between prominent families in early 15[th] century Florence.

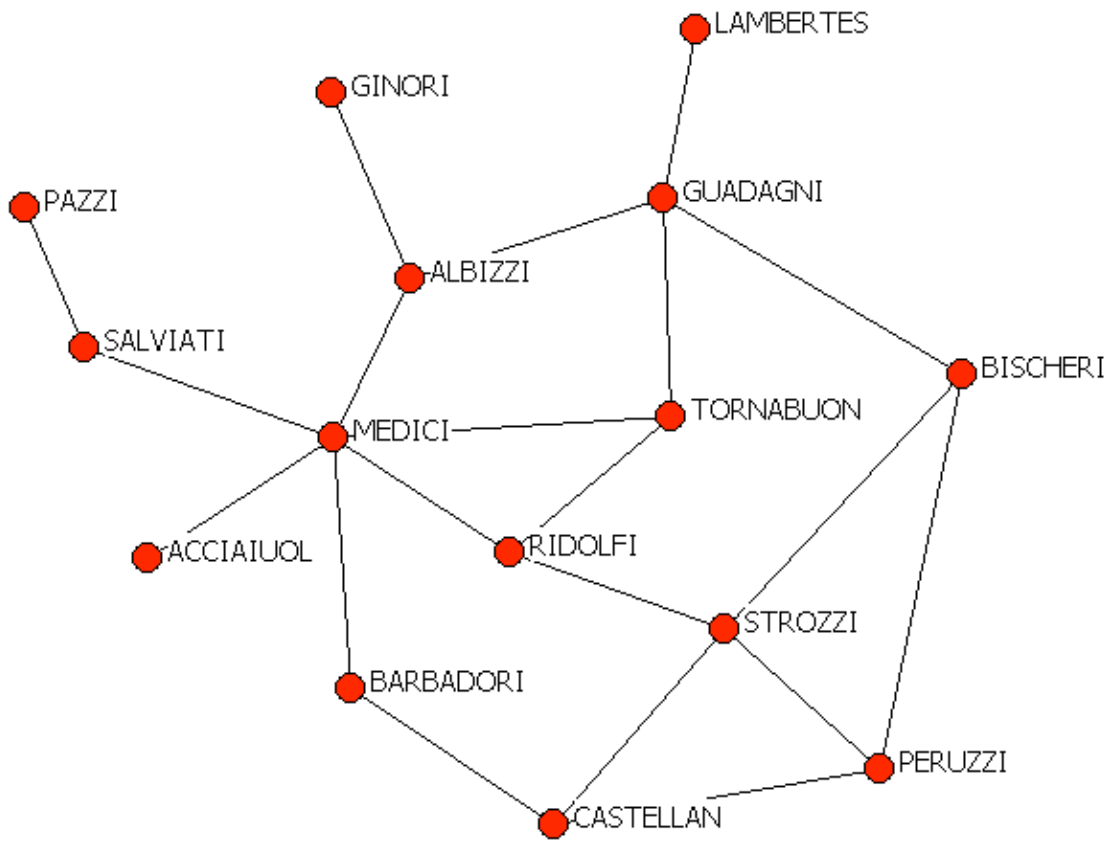

Fig. 5. Proportion of conserved to non-conserved proteins in the *E. coli* PIN detected by the ranking of proteins according to different centrality measures studied here.

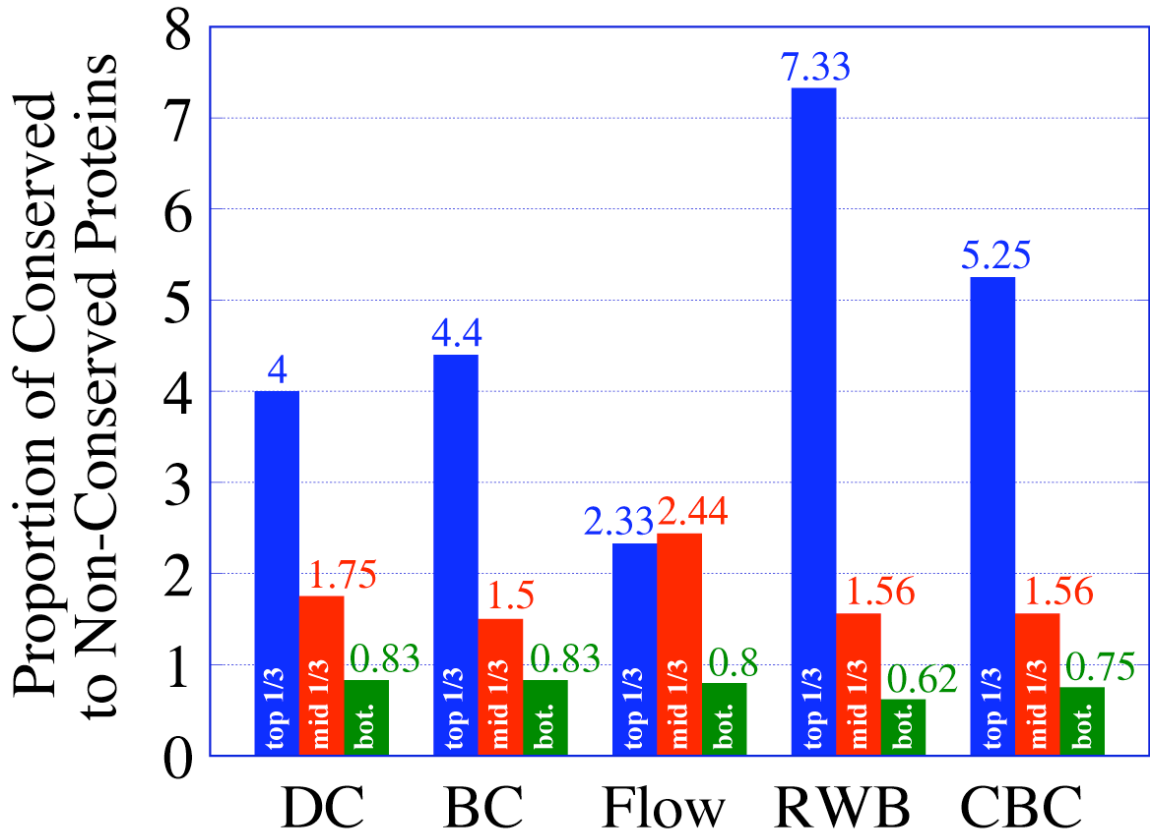

Fig. 6. Average number of genomes in which one of the top 20 proteins according to the ranking introduced by the centrality measures appears.

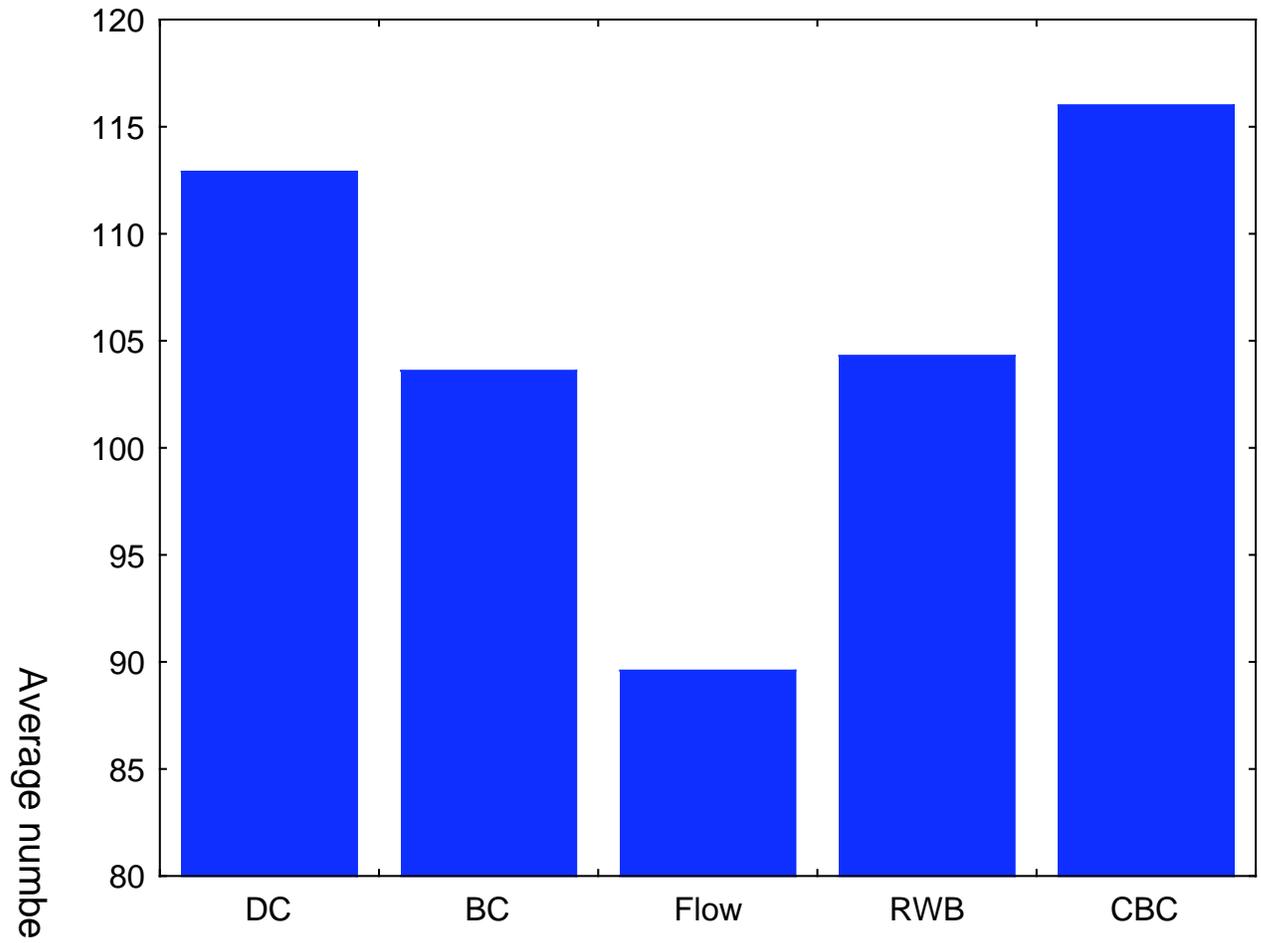

Fig. 7. Number of genomes in which a protein appears according to the ranking introduced by the different centrality measures studied here.

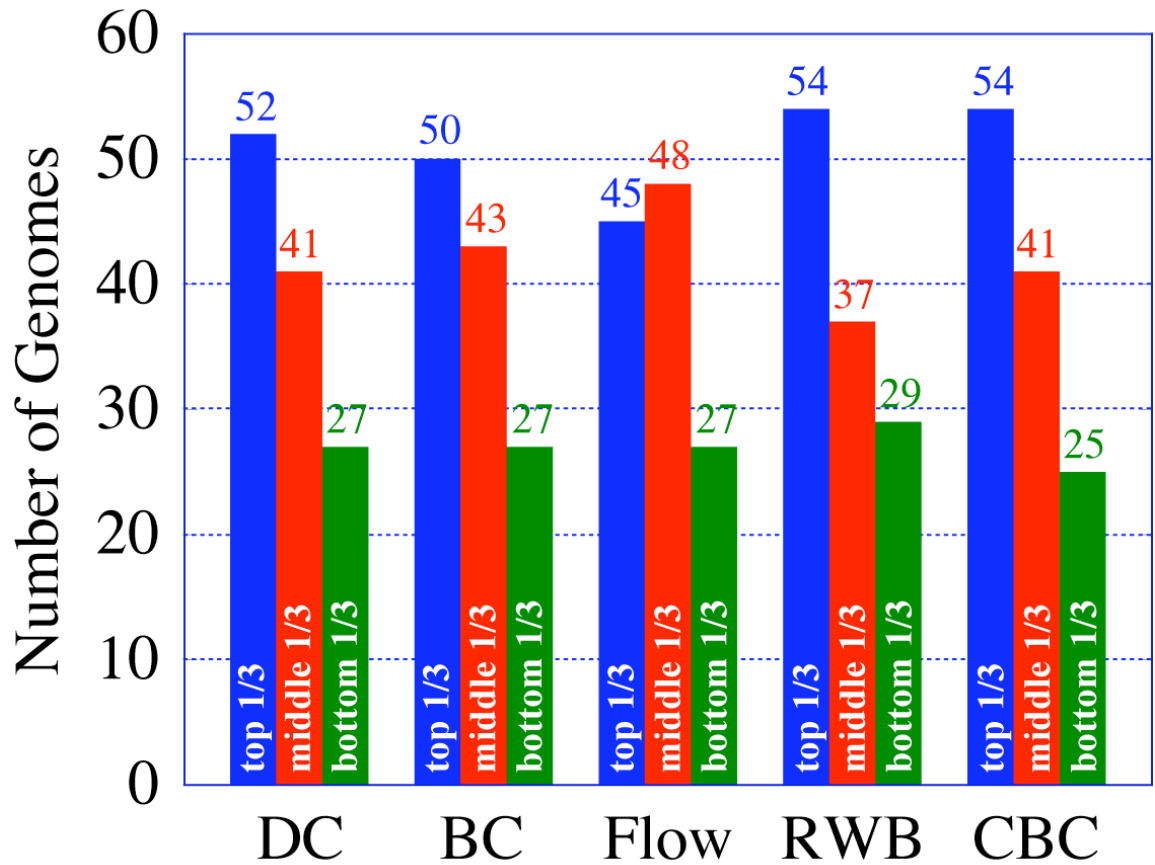